\documentclass[useAMS,usenatbib]{mn2e}

\usepackage{times}
\usepackage{graphicx}

\def\pcm3{{\rm\thinspace cm^{-3}}}

\def\contcaption{\@conttrue\SFB@caption\@captype}

\def\n_h{{\rm n_{H}}}

\def\NH1{{$N_{\rm HI}~$}}

\def\ga{{\rm\thinspace gauss}}

\def\approxlt{\mathrel{\hbox{\rlap{\lower .5ex \hbox {$\sim$}}
        \raise .15 ex \hbox{$<$}}}}
\def\approxgt{\mathrel{\hbox{\rlap{\lower .5ex \hbox {$\sim$}}
        \raise .15 ex \hbox{$>$}}}}

\def\la{\mathrel{\hbox{\rlap{\hbox{\lower4pt\hbox{$\sim$}}}\hbox{$<$}}}}
\def\ga{\mathrel{\hbox{\rlap{\hbox{\lower4pt\hbox{$\sim$}}}\hbox{$>$}}}}
\newbox\grsign \setbox\grsign=\hbox{$>$} \newdimen\grdimen
\grdimen=\ht\grsign
\newbox\simlessbox \newbox\simgreatbox \newbox\simpropbox
\setbox\simgreatbox=\hbox{\raise.5ex\hbox{$>$}\llap
     {\lower.5ex\hbox{$\sim$}}}\ht1=\grdimen\dp1=0pt
\setbox\simlessbox=\hbox{\raise.5ex\hbox{$<$}\llap
     {\lower.5ex\hbox{$\sim$}}}\ht2=\grdimen\dp2=0pt
\setbox\simpropbox=\hbox{\raise.5ex\hbox{$\propto$}\llap
     {\lower.5ex\hbox{$\sim$}}}\ht2=\grdimen\dp2=0pt
\def\simgreat{\mathrel{\copy\simgreatbox}}
\def\simless{\mathrel{\copy\simlessbox}}

\title[Two new young, wide, DA+DAH double-degenerates]{Two new young, wide, magnetic + non-magnetic double-degenerate binary systems.\thanks{Based on observations made with ESO Telescopes at the La Silla or Paranal Observatories under programme ID 084.D-1097}}
\author[Dobbie, Baxter, K\"ulebi, Parker, Koester, Jordan, Lodieu \& Euchner]{P. D. Dobbie$^{1}$\thanks{E-mail:
pdd@aao.gov.au}, R. Baxter$^{2}$, B. K\"ulebi$^{3}$, Q. A. Parker$^{1,2}$, D. Koester$^{4}$, S. Jordan$^{5}$, \newauthor
 N. Lodieu$^{6,7}$, F. Euchner$^{8}$ \\
$^{1}$Australian Astronomical Observatory, PO Box 296, Epping, NSW, 1710, Australia \\
$^{2}$Dept. of Physics \& Astronomy, Macquarie University, NSW, 2109, Australia\\
$^{3}$Institut de Ci\`{e}ncies de l$^{\prime}$Espai (CSIC-IEEC), Facultat de Ci\`{e}ncies, Campus UAB, Torre C5-parell, 2$^{\rm a}$  planta, 08193 Bellaterra, Spain \\
$^{4}$Institut f\"ur Theoretische Physik und Astrophysik, Christian-Albrechts-Universit\"at, Kiel, Germany \\
$^{5}$Astronomisches Rechen-Institut, Zentrum f\"ur Astronomieder Universit\"at Heidelberg, M\"onchhofstr. 12–14, D-69120 Heidelberg, Germany \\
$^{6}$Instituto de Astrof\'isica de Canarias, V\'ia L\'actea s/n, E-38200 La Laguna, Tenerife, Spain \\
$^{7}$Departmento de Astrof\'isica, Universidad de La Laguna, E-38205 La Laguna, Tenerife, Spain \\
$^{8}$Swiss Seismological Service, ETH Zurich, Schafmattstrasse 30, HPP P3, 8093 Zurich, Switzerland
}
\begin{document}

\date{Accepted 2011 November 29. Received 2011 November 28; in original form 2011 September 4}

\pagerange{\pageref{firstpage}--\pageref{lastpage}} \pubyear{2009}

\maketitle

\label{firstpage}

\begin{abstract}

We report the discovery of two, new, rare, wide, double-degenerate binaries that each contain a magnetic and a non-magnetic star. The components of SDSS\,J092646.88+132134.5 + J092647.00+132138.4 and SDSS\,J150746.48+521002.1 + J150746.80+520958.0 have angular separations of only 4.6 arcsec (a$\sim$650AU) and 5.1 arcsec (a$\sim$750AU),
respectively. They also appear to share common proper motions. Follow-up optical spectroscopy reveals each system to consist of
a  DA and a H-rich high-field magnetic white dwarf (HFMWD). Our measurements of the effective temperatures and the
surface gravities of the DA components reveal both to have larger masses than are typical of field white dwarfs. 
By assuming that these degenerates have evolved essentially as single stars, due to their wide orbital separations, we use them to place limits on the total ages of
our stellar systems. These argue that in each case the HFMWD is probably associated with an early type progenitor ($M_{\rm init}$$>$2M$_{\odot}$). 
We find that the cooling time of SDSS\,J150746.80+520958.0 (DAH) is somewhat lower than might be expected had it followed the evolutionary path of a typical single star. This mild discord is in the same sense as that observed for two of the small number of other HFMWDs for which progenitor mass estimates have been made, RE\,J0317-853 and EG\,59. The mass of the other DAH, SDSS\,J092646.88+132134.5, appears to be smaller than expected on the basis
of single star evolution. If this object was/is a member of a hierarchical triple system it may have experienced greater mass 
loss during an earlier phase of its life as a result of it having a close companion. The large uncertainties on our estimates of the parameters of the HFMWDs suggest a larger sample 
of these objects is required to firmly identify any trends in their inferred cooling times and progenitor masses. This should shed further light on their formation and the impact magnetic fields have
on the late stages of stellar evolution. To serve as a starting point, we highlight two further candidate young, wide 
magnetic + non-magnetic double-degenerate systems within SDSS, CBS\,229 and SDSS\,J074853.07+302543.5 + J074852.95+302543.4, which 
should be subjected to detailed (resolved) spectroscopic followed-up studies.

\end{abstract}


\section{Introduction}

\begin{table*}
\begin{minipage}{171mm}
\begin{center}
\caption{A summary of the photometric and astrometric properties of the components of our two putative wide double-degenerate systems.}

\begin{tabular}{lcccccccc}
\hline
\multicolumn{1}{c}{IAU} & Name  & $u$ & $g$ & $r$ & $i$ & $z$ &  $\mu_{\alpha}$cos $\delta$ & $\mu_{\delta}$ \\ 

  & & \multicolumn{5}{c}{SDSS} &  \multicolumn{2}{c}{(mas yr$^{-1}$)} \\
%  & ($^{\rm h}$ $^{\rm m}$ $^{\rm s}$$_{.}$) & ($^{\circ}$ ' ``$_{.}$) \\
\hline

SDSS\,J092646.88+132134.5 & DAH1  &18.46$\pm$0.02 & 18.34$\pm$0.02 & 18.39$\pm$0.01 & 18.50$\pm$0.02 & 18.60$\pm$0.03 & -8.6$\pm$6.9 & -77.2$\pm$9.6 \\
SDSS\,J092647.00+132138.4 & DA1   &18.74$\pm$0.03 & 18.40$\pm$0.03 & 18.46$\pm$0.05 & 18.60$\pm$0.04 & 18.79$\pm$0.03 &-11.6$\pm$6.9 & -65.3$\pm$9.6 \\ \\

SDSS\,J150746.48+521002.1 & DA2   &17.14$\pm$0.02 & 16.91$\pm$0.03 & 17.29$\pm$0.01 & 17.55$\pm$0.02 & 17.84$\pm$0.02 &-30.3$\pm$4.9 & +12.7$\pm$5.8 \\
SDSS\,J150746.80+520958.0 & DAH2  &17.98$\pm$0.03 & 17.76$\pm$0.03 & 18.06$\pm$0.01 & 18.33$\pm$0.02 & 18.55$\pm$0.03 &-31.0$\pm$4.9 & +13.1$\pm$5.8 \\

\hline
\label{phot}
\end{tabular}
\end{center}
\end{minipage}
\end{table*}

A non-negligible proportion of white dwarfs appear to possess substantial magnetic fields, with strengths typically $>$1MG.  A number of studies have determined that they represent between $\sim$5-15 per cent of the white dwarf population yet their origins remain quite unclear \citep{angel81, liebert03, kawka07,kulebi09}. These are often referred to as the high field magnetic white dwarfs \citep[HFMWDs, e.g.][]{wickram05}. While the mass distribution of field white dwarfs is found to be strongly peaked around 0.6M$_{\odot}$ \citep[e.g.][]{liebert05a,koester09}, the mass distribution of the HFMWDs is flatter and skewed towards higher masses, $M$$\sim$0.9M$_{\odot}$ \citep[e.g.][]{liebert03}. Three of the ten ultramassive ($M$$>$1.1M$_{\odot}$) white dwarfs identified in the extreme ultraviolet surveys appear to be HFMWDs \citep{vennes99}. 

At present, there are two principle theories regarding their formation. In the ``fossil field'' hypothesis the HFMWDs are the descendents of the Ap + Bp stars, a magnetic, chemically peculiar subset of objects with spectral types ranging from late-B to early-F \citep{angel81}. This is in accord with the similar magnetic fluxes of the HFMWDs and the Ap + Bp stars and with the predicted long decay times of the fields in these objects. Moreover, the higher average mass of the HFMWDs is explained naturally here as a result of the form of the stellar initial-final mass relation, a positive correlation between the main sequence masses of stars and their white dwarf remnant masses \citep[e.g.][]{weidemann00}. However, in light of more recent results, the proportion of late-B to early-F stars that can be classified as Ap + Bp may be too low by a factor of 2-3 to be consistent with the larger revised estimates of the percentage of HWMWDs in the general white dwarf population \citep[e.g.][]{kawka03}. To alleviate this apparent shortfall in progenitors, it is required that $\sim$40 per cent of stars with M$>$4.5M$_{\odot}$ also evolve to become HFMWDs \citep[e.g.][]{wickram05}.

Alternatively, \cite{tout08} have proposed that the magnetic fields of HFMWDs are generated by differential rotation within the common envelope gas which engulfs a primordial close binary system when the primary star expands to giant dimensions and overfills its Roche Lobe. An isolated HFMWD is predicted to form if the cores of the components merge before this envelope is dispersed. However, if the gas is removed prior to this, the outcome is instead expected to be a magnetic-cataclysmic variable. This hypothesis can account for the puzzling lack of detached HFMWD + late-type star binary systems that has emerged from the Sloan Digital Sky Survey \citep[SDSS; e.g.][]{liebert05c}. It might also explain the reported discrepancies in the cooling times of a small number of the HFMWDs where it has been possible to test them against evolutionary models for single stars \citep[e.g. RE J0317-835 and EG59, ][]{barstow95, ferrario97, claver01, casewell09}.
 
\begin{figure*}
\includegraphics[angle=270,width=15cm]{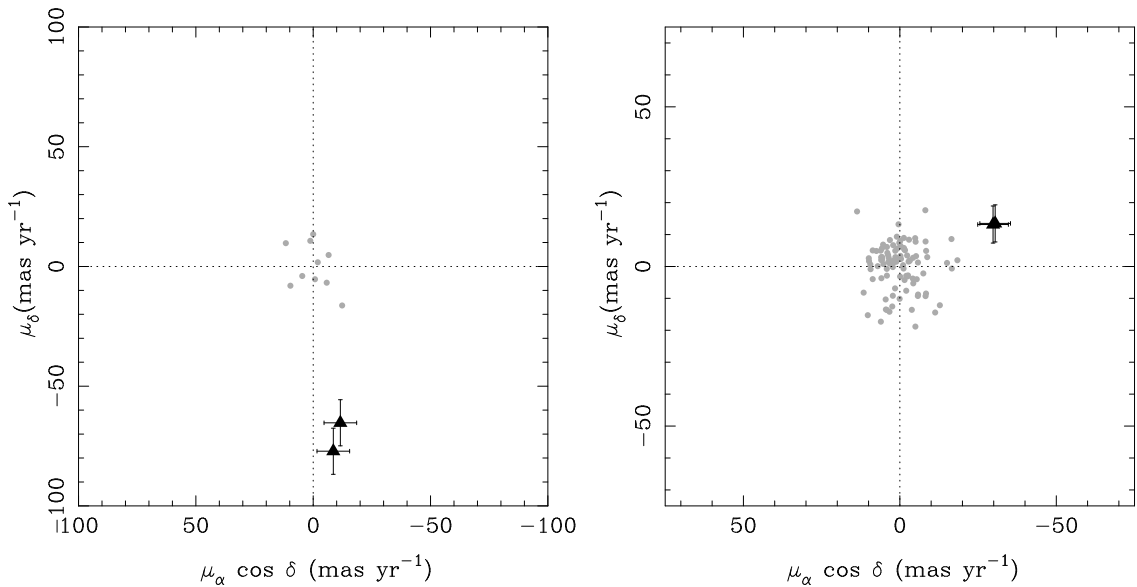}
\caption{A vector point diagram of the relative proper motions of the SDSS\,J092647.00+132138.4 + J092646.88+132134.5 system (left) and the SDSS\,J150746.48+521002.1 + J150746.80+520958.0 system (right).
The reference stars used in the construction of the linear transform are also shown (light grey points). The similarities between the proper motions of the components of these putative binary systems is clear to see (solid triangles).}
\label{VPD}
\end{figure*}

The identification of new HFMWDs where there is the opportunity to set constraints on their masses, cooling times and the ages of their parent populations can further address the questions regarding their origins. Traditionally, white dwarf members of open clusters are used since their masses can be constrained from their observed fluxes using a mass-radius relation \citep[e.g.][]{casewell09}. Subsequently, their cooling times can be derived and compared to those of the non-magnetic degenerate members for which progenitor masses can be estimated straightforwardly. Unfortunately, less than a handful of HFMWDs have been found in open clusters to date. These have tended to be rather faint due to their substantial distances \citep[e.g. NGC\,6819-8, $V$$\approx$23.0, ][]{kalirai08} and thus not be particularly amenable to detailed follow-up study. 

An alternative approach focuses on field HFMWDs in nearby, wide, double-degenerate systems where the components are sufficiently far apart to have evolved essentially as separate entities yet the system age and distance can be determined from the non-magnetic companion star \citep[e.g.][]{girven10}. However, only two of these spatially resolved binaries have been identified to date. RE\,J0317-853 \citep{barstow95}, which resides at d$\sim$30pc \citep{kulebi10}, was discovered in the course of the extreme-ultraviolet all sky surveys undertaken with the {\it R\"{o}ntgensatellit} \citep[e.g.][]{pye95} and the {\it Extreme Ultraviolet Explorer} \citep[e.g.][]{bowyer96} satellites. It consists of a common proper motion pairing of a $\sim$0.85M$_{\odot}$ DA and an ultramassive HFMWD with a field strength of $B$$\sim$450MG \citep{ferrario97,burleigh99} that are separated on the sky by 7 arcsec. The more recently discovered common proper motion system PG\,1258+593  +  SDSS\,J130033.48+590407.0 \citep{girven10} contains a pair of near equal mass ($M$$\approx$0.54M$_{\odot}$) hydrogen rich white dwarfs that are separated on sky by 16 arcsec.  The HFMWD component, which has a field strength of $B$$\sim$6MG, is the cooler of the pair. In addition, three unresolved double-degenerate systems containing a magnetic and a non-magnetic object are also currently listed in the refereed literature. LB\,11146 \citep{liebert93} is known to be a physically close system \citep{nelan07} while the orbital separations of the white dwarfs in RE\,J1439+75 \citep{vennes99} and G\,62-46 \citep{bergeron93} might also be relatively small. 

With several large area charged-coupled device (CCD) imaging surveys such as SDSS \citep{york00}, VST ATLAS (http://www.astro.dur.ac.uk/Cosmology/vstatlas/) and SkyMapper \citep{keller07} having recently been completed or soon to be undertaken, the prospects for unearthing more young, wide, magnetic + non-magnetic white dwarf binaries are excellent. Here we report the discovery and confirmation of two further examples of such systems, SDSS\,J092646.88+132134.5 + J092647.00+132138.4 (hereafter, System 1) and SDSS\,J150746.48+521002.1 + J150746.80+520958.0 (hereafter, System 2). In the following sections we briefly describe our broader survey for wide double-degenerate binaries and demonstrate that the components of these two new pairings share common proper motions. We perform a spectroscopic analysis of the two white dwarfs in each system to assess their masses and cooling times. Subsequently, we examine the objects in the context of canonical single star evolutionary theory and place constraints on the nature of the progenitors of the HFMWDs. Additionally, we search for evidence that these white dwarfs have more exotic formation histories.  We finish by highlighting two further candidate wide, magnetic + non-magnetic double-degenerate binaries which maybe suitable for this type of analysis. These can serve as a starting point for the construction of an enlarged sample of these systems that will be neccessary to better understand the formation of HFMWDs.

\section[]{Discovery of young, wide, magnetic + non-magnetic white dwarf binaries.}

\subsection{Imaging search for wide, double-degenerate systems.}
\label{survey}

We have been conducting a survey for young, wide, double-degenerates using imaging and spectroscopic data from the SDSS. The full details of this study will be described in a forthcoming paper (Baxter et al. in prep) but we provide a brief outline of our approach here so that this result can be placed into context. We selected from DR7 (the 7th SDSS data release) all point sources flagged as photometrically clean with $r$$\le$20.0, $u$-$g$$\le$0.5, $g$-$r$$\le$0.0 and $r$-$i$$<$0.0 (corresponding to white dwarfs with $T_{\rm eff}$$\simgreat$9000K \citealp[e.g.][]{eisenstein06}) which have another object satisfying these colour-magnitude criteria within 30 arcsec. The resulting candidate systems were visually inspected using the SDSS finder chart tool to weed out a number of spurious pairings (e.g. blue point-like detections within resolved galaxies). This procedure led to the identification of 52 candidate systems, including the previously known double-degenerate binaries, PG\,0922+162 \citep{finley97} and HS\,2240+1234 \citep{jordan98}. The associated SDSS spectroscopy, which is available for 21 objects in 19 of these pairs, reveals 17 DAs, 1 DB and 3 DCs but no quasars or subdwarfs suggesting that contamination levels in this sample are low. We have obtained long-slit spectroscopic follow-up data for 13 additional candidate binaries to date and all have turned out to be comprised of white dwarfs, confirming the low level of contamination. 

System 1 and System 2 are two of these 13 pairings which both appear to harbour a HFMWD. Their components are separated on the sky by only 4.35 arcsec and 5.05 arcsec, respectively. Their $u$, $g$, $r$, $i$ and $z$ magnitudes are listed in Table~\ref{phot}. To ascertain the likelihood of chance alignments we have used Equation 1 \citep[][]{struve52}, under the assumption of a random on-sky distribution of objects, where $N$ is the number of sources satisfying the photometric selection criteria in area $A$ (square degrees) and $\rho$ is the maximum projected separation (degrees).

\begin{eqnarray}
n(\le\rho)= N (N-1) \pi \rho^{2} / 2 A
\end{eqnarray}

We estimate $n$(4 arcsec$<$$\rho$$\le$6 arcsec)$\sim$0.3 for $N$=36231 and $A$=11663 square degrees. In the course of our 
survey we have unearthed a total of 10 pairs of objects with separations in this range, suggesting that the probability of 
any one being a mere chance alignment is P$\sim$0.03.

\subsection{Astrometric follow-up and proper motions}

To probe the reality of the putative associations between the two sources in each of these pairs, we have examined their relative proper motions. We note that \cite{girven10} used proper motions to confirm the association of PG\,1258+593  +  SDSS\,J130033.48+590407.0 in their recent exploration of the bottom end of the IFMR. In principle, measurements of this nature can be obtained from one of a number of online databases featuring imaging of widely separated epochs e.g. the United States Naval Observatory B1.0 catalogue \citep[USNO-B1.0;][]{monet03}, SDSS \citep[e.g.][]{munn04}, the SuperCOSMOS Sky Survey \citep[SSS;][]{hambly01a} and the extended Position and Proper Motion catalogue \citep[PPMXL;][]{roeser10}. However, these resources rely heavily on comparatively low spatial resolution photographic plate exposures. Examination of the values reported in these catalogues indicates that the small separations of the components of our candidate binaries have compromised their accuracy. For example, the astrometry reported in the PPMXL catalog suggests that the relative positions of the objects in these two pairs should have changed by several arc seconds in the $\sim$50 years between the imaging of the first Palomar Sky Survey and SDSS. However, visual inspection of these data reveals no discernable difference. Therefore, we have opted to perform our own astrometric measurements of System 1 and System 2. For the first system we have adopted the $g$ band data from SDSS (2006/01/06) and a $B$ band acquisition frame (2010/02/06) from our spectroscopic follow-up program on the European Southern Observatory's (ESO) Very Large Telescope (VLT) as first and second epoch images respectively. We have used \citep[{\tt SExtractor} ][]{bertin96} to determine the positions of $\sim$10 unblended, stellar-like objects lying in close proximity to the candidate white dwarfs. As no suitable second CCD image of System 2 is available to us we have resorted to using data from the first Palomar Sky Survey (Plate 2376 observed on 1954/06/28) and the SDSS $r$ band imaging (2002/06/09) as our first and second epoch images of this system, respectively. Here we have determined the positions of unsaturated stellar-like objects lying within a few arcminutes of the candidate white dwarfs. In detecting these stars and determining their centroids in the photographic data, no smoothing filter was applied to the image (to minimise blending of the stellar profiles) and only those pixels substantially above the background ($\simgreat$5$\sigma$) were included in the calculations.

\begin{figure}
\includegraphics[angle=270,width=\linewidth]{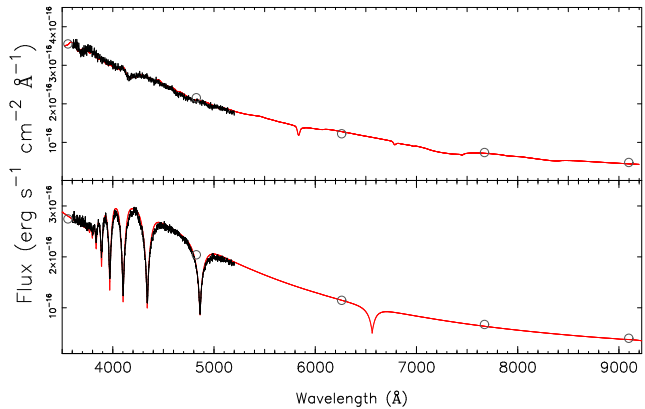}
\caption{VLT + FORS2 spectroscopy of SDSS\,J092646.88+132134.5  (upper) and SDSS\,J092647.00+132138.4 (lower) with SDSS $u$, $g$, $r$, $i$ and $z$ 
fluxes (open grey circles) and synthetic spectra (red line) overplotted. The spectrum of SDSS\,J092647.00+132138.4 is clearly that of an H-rich white dwarf, while that
of SDSS\,J092646.88+132134.5 displays several weaker features which are consistent with magnetically broadened and shifted Balmer lines. 
}
\label{specs1}
\end{figure}

For each pair of objects we have cross matched the lists of reference star positions using the {\tt{STARLINK TOPCAT}} software. 
Subsequently, we have employed routines in the {\tt{STARLINK SLALIB}} library to construct six co-efficient linear 
transforms between the two images of the putative systems, where $>$3$\sigma$ outliers were iteratively clipped from 
the fits. The proper motions, in pixels, were determined by taking the differences between the observed and calculated 
locations of candidates in the 2nd epoch imaging. These were then converted into milli-arcseconds per year in right 
ascension and declination using the world co-ordinate systems of the 1st epoch datasets and dividing by the time baseline between the two observations, $\sim$4.08\,yr for System 1 and  $\sim$47.95\,yr for System 2 (Table~\ref{phot}). The uncertainties on these measurements were estimated from the dispersion
observed in the (assumed near-zero) proper motions of stars of comparable brightness surrounding each system. The relative 
proper motion vector point diagrams for the objects (solid triangles) are shown in 
Figure~\ref{VPD}. 

\begin{figure}
\includegraphics[angle=270,width=\linewidth]{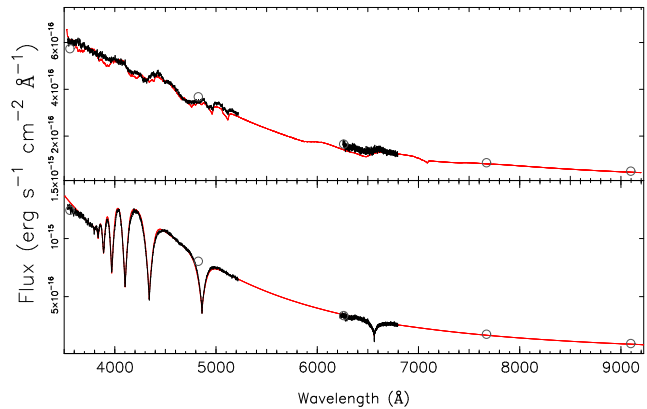}
\caption{WHT + ISIS spectroscopy of SDSS\,J150746.48+521002.1 (lower) and SDSS\,J150746.80+520958.0 (upper) with SDSS $u$, $g$, $r$,
$i$ and $z$ fluxes overplotted (open grey circles). The spectrum of SDSS\,J150746.48+521002.1 is clearly that of an H-rich white dwarf,
while that of SDSS\,J150746.80+520958.0 displays several weaker features which are consistent with magnetically broadened and shifted 
Balmer lines.
}
\label{specs2}
\end{figure}

\begin{table*}
\begin{minipage}{135mm}
\begin{center}
\caption{Effective temperatures, surface gravities, predicted absolute $r$ magnitudes,
 distance modulii, masses and cooling times for the DA components of the two wide binary systems.}
\label{wdmass}
\begin{tabular}{clcrccc}
\hline
SDSS & $T_{\rm eff}$$^{*}$ & log~$g$$^{*}$ & M$_{r}$ & $r$-M$_{r}$ & M(M$_{\odot}$)  & $\tau_{c}$ (Myr) \\ 
\hline
 \multicolumn{1}{l}{J092647.00+132138.4} & $10482^{+47}_{-47}$ & $8.54^{+0.03}_{-0.03}$ &$12.57^{+0.17\ddagger}_{-0.17}$ & 5.89$\pm$0.18$^{\ddagger}$ & 0.79$\pm$0.06$^{\ddagger}$ & 833$^{+123\ddagger}_{-123}$\\ \\
\multicolumn{1}{l}{J150746.48+521002.1} & $17622^{+99}_{-94}$ & $8.13^{+0.02}_{-0.02}$ & $11.38^{+0.11}_{-0.11}$ & 5.91$\pm$0.11 & 0.70$\pm$0.04 & 147$^{+24}_{-21}$ \\                    
\hline
\label{temps}
\end{tabular}
\end{center}
$^{*}$ Formal fit errors. \\
$^{\ddagger}$ Adjusted to account for spectroscopic overestimate of mass.
\end{minipage}
\end{table*}

Inspection of these plots reveals that our proper motion measurements of the components of the pairs are significant at $>$3$\sigma$ and consistent with each other within their 1$\sigma$ uncertainties. For each case we have estimated the probability that this similarity could have occurred merely by chance. We have selected all spectroscopically confirmed white dwarfs within 10$^{\circ}$ of these systems from the SDSS DR4 white dwarf catalogue \citep{eisenstein06} which meet the photometric selection criteria outlined in Section~\ref{survey}. We have cross-correlated these with the SSS database to obtain their proper motions. After cleaning these samples for objects with poorly constrained astrometry (e.g. $\chi^{2}$$>$3 or flagged as bad and/or proper motion uncertainties $>$7.5mas~yr$^{-1}$), we are left with 59 and 159 white dwarfs in the general directions of System 1 and System 2, respectively. An examination of this astrometry reveals that none and seven sources in these samples have proper motions that could be deemed as consistent (within 2$\sigma$ of the mean proper motion of the components of the putative system) with System 1 and System 2, respectively. Therefore we estimate the probabilities that the proper motions of these objects are similar by chance are less than 0.02 and 0.05 respectively. Combined with the likelihoods of chance projected angular proximity we find the probabilities of these two systems merely being optical doubles are less than P$\sim$0.0006 and P$\sim$0.0015 giving a potent argument in favour of their association.

\section[]{Spectroscopic analysis}

\subsection{Follow-up optical spectroscopy of the binary components.}
\label{followup}

We have obtained optical spectroscopy of System 1 in visitor mode with the ESO Very Large Telescope and the Focal Reducer and low dispersion Spectrograph (FORS2). 
A full description of the FORS2 instrument may be found on the ESO webpages\footnote{http://www.eso.org/instruments/fors2/}. These observations (1$\times$360s and 1$\times$500s exposures) were conducted on the night of 2010 February 6 when the seeing was good but there 
was some cirrus scattered 
across the sky. All data were acquired using the 2$\times$2 binning mode of the $E2V$ CCD, the 600B+24 grism and a 1.3 arcsec 
slit which gives a notional resolution of $\lambda$/$\Delta$$\lambda$$\sim$600. 
The components of System 2 were observed with the William Herschel Telescope (WHT) 
and the double-armed Intermediate dispersion Spectrograph and Imaging System (ISIS) on the night of 2008 July 24. These observations were conducted when the sky was photometric, with seeing $\sim$0.6-0.9 arcsec. The spectrograph was configured with a 1.0 arcsec slit and with the R300B ($\lambda$/$\delta\lambda$$\approx$2000) and the R1200R ($\lambda$/$\delta\lambda$$\approx$10000) gratings on the blue and red arms, respectively. Three 1800s exposures covering the two wavelength ranges, $\lambda$$\approx$3600-5200\AA\ and 6200-7000\AA, were obtained simultaneously. 

The CCD frames were debiased and flat fielded using the IRAF procedure CCDPROC. Cosmic ray hits were removed using the routine LACOS SPEC (van Dokkum 2001). Subsequently, the spectra were extracted using the APEXTRACT package and wavelength calibrated by comparison with a CuAr+CuNe arc spectrum taken immediately before and after the science exposures (ISIS) or with a He+HgCd arc spectrum obtained within a few hours of the science frames (FORS2). The removal of remaining instrument signature from the science data was undertaken using observations of the bright DC white dwarfs WD1918+386 (ISIS) and LHS2333 (FORS2). The spectra of the components of System 1 and System 2 are shown in Figures~\ref{specs1} and \ref{specs2}.

\subsection[]{Effective temperatures and surface gravities}
\label{tandg}

The optical energy distributions of SDSS\,J092647.00+132138.4 (hereafter, DA1) and SDSS\,J150746.48+521002.1 (hereafter, DA2) each display broadened H-Balmer line series 
and these objects are unmistakably hydrogen rich white dwarfs. In contrast, no prominent features at the notional wavelengths of 
the lines of either {\tt HI} or {\tt HeI} are observed in the spectra of  SDSS\,J092646.88$+$132134.5 (hereafter, DAH1) and SDSS\,J150746.80$+$520958.0 (hereafter, DAH2). However, neither dataset is completely 
smooth like the spectrum of a DC white dwarf and instead they display a number of shallow depressions across the observed wavelength range 
reminiscent of a H-rich HFMWD such as SDSS\,J172045.37+561214.9 \citep{gaensicke02}. 

To determine the effective temperatures and the surface gravities of DA1 and DA2 we have compared 
the observed Balmer lines, H-8 to H-$\beta$, to a grid of synthetic profiles \citep[e.g.][]{bergeron92}. These are based on recent versions of the plane-parallel, hydrostatic,
local thermodynamic equilibrium (LTE) atmosphere and spectral synthesis codes {\tt ATM} and {\tt SYN} \citep[e.g.][]{koester10}, which include an 
updated physical treatment of the Stark broadening of {\tt HI} lines \citep{tremblay09}. The fitting procedure has been performed with the spectral
analysis package {\tt XSPEC} \citep{shafer91}.  {\tt XSPEC} folds a model through the instrument response before comparing the result to the data by 
means of a $\chi^{2}-$statistic. The best model representation of the data is found by incrementing free grid parameters in small steps, linearly 
interpolating between points in the grid, until the value of $\chi^{2}$ is minimised.  Formal errors in the $T_{\rm eff}$s and log $g$s are calculated 
by stepping the parameter in question away from its optimum value and redetermining minimum $\chi^{2}$ until the difference between this and the true
minimum $\chi^{2}$ corresponds to $1\sigma$ for a given number of free model parameters \citep[e.g. see][]{lampton76}. It is important to be aware that these errors do not take into account shortcomings in the models or systematic issues affecting the data (e.g. flaws in the flat fielding) so 
undoubtedly underestimate the true uncertainties in the parameters. Therefore in our subsequent analysis we assume more realistic levels of uncertainty of 2.3 per cent and 0.07dex in effective temperature and surface gravity, respectively \citep[e.g. see][]{napiwotzki99}. The results of our line fitting are shown in Table~\ref{temps}. Synthetic colours for H-rich white dwarfs at these effective temperatures and surface gravities \citep[e.g.][]{holberg06} appear to be broadly consistent with the SDSS photometry for DA1 and DA2 (Table~\ref{phot}). 

\begin{table*}
\begin{minipage}{138mm}
\begin{center}
\caption{Effective temperatures, field strengths and geometries, masses and cooling times for the magnetic components of the two binary 
systems.}
\label{magnetic}
\begin{tabular}{ccccccc}
\hline
SDSS &  $T_{\rm eff}$	& $B_{\rm dip}$(MG) &	$z_{\rm off}({\rm R}_{\rm WD})$	& inclination (${}^\circ$) & $M$(M$_{\odot}$) & $\tau_{c}$ (Myr) \\
\hline
J092646.88$+$132134.5 & 9500$\pm$500	& 210$\pm$25.1   & -0.09$\pm$0.01 & 21.8$\pm$7.9 & 0.62$\pm$0.10 & 726$^{+140}_{-107}$\\
J150746.80$+$520958.0 &	18000$\pm$1000  &  65.2$\pm$0.3 &  -0.39$\pm$0.03 & 36.4$\pm$4.1 & 0.99$\pm$0.05 & 321$^{+47}_{-40}$ \\
\hline
\label{magnetic}
\end{tabular}
\end{center}
\end{minipage}
%\label{temps}
\end{table*}

The energy distributions of the DAH components DAH1 and DAH2 have been compared to a grid of model spectra for HFMWDs. These were calculated with a radiative transfer code for magnetised, high gravity atmospheres. The code calculates both theoretical flux (Stokes $I$) and polarization (Stokes $V$) spectra, for a given temperature and pressure structure ($T_{\rm eff}$, $\log g$) and a specific magnetic field vector with respect to the line of sight and the normal on the surface of the star \citep[see][]{jordan92,jordan93}. However, as no polarization information is available from our datasets, our analysis has been limited to the flux spectra (Stokes parameter $I$). For computational efficiency, we have made use of our existing three-dimensional grid of synthetic spectra. This grid spans the effective temperature range $7000\,{\rm K}\le T_{\rm eff} \le 50000\,{\rm K}$ in 14 steps, treats magnetic field strengths between $1\,{\rm MG}\le B\le 1.2\,{\rm GG}$ in 1200 steps and has 17 different values of the field direction $\psi$ relative to the line of sight, as the independent variables (9 entries, equally spaced in $\cos \psi$). All spectra were calculated for a surface gravity of log $g$=8 \citep[see][]{kulebi09}. Limb darkening is accounted for by a simple linear scaling law \citep[see][]{euchner02}.

The magnetic field geometry of the DAHs has been determined using a modified version of the code developed by \citet{euchner02}. This code 
calculates the total flux distribution for an arbitrary magnetic field topology by adding up appropriately weighted synthetic spectra for a 
large number of surface elements and then evaluates the goodness of fit. For this work we have used a simple model for the magnetic geometry, 
namely a dipole with an offset on its polar axis. This is the most parameter efficient way of generating non-dipolar geometries and is 
representative of dipole plus quadrupole configurations. Moreover, these models are ubiquitous in the diagnosis of single phase HFMWD spectra 
\citep{kulebi09}. The resulting magnetic parameters of our fits are summarized in Table~\ref{magnetic}. As explained in \citet{euchner02}, these values might not be unique and it is possible to fit similar spectra with different models. However this analysis is fully satisfactory in the context of this work since we are interested in a good spectral model of a given object for determining the temperature. We find that the magnetic field geometries of these two HFMWDs are quite different. While the field of DAH1 does not deviate significantly from dipolarity, in the case of DAH2 the offset parameter is quite large and the magnetic field distribution used in the construction of the fitted spectrum departs substantially from a simple dipole with field strengths of between 45-317MG.

After obtaining a satisfactory representation of the magnetic field structures, the effective temperature of each star has been assessed from the line strengths and the continuum \citep[e.g.][]{gaensicke02}. These are also shown in Table~\ref{magnetic}. While in some cases the magnetic analysis can be hindered by the lack of a realistic and easily applicable theory for the simultaneous impact of Stark and Zeeman effect on the spectral lines, this is not a problem for the field strengths relevant to this work ($B>50$\,MG). 

\subsection{White dwarf masses and cooling times}

We have used our measurements of the effective temperatures and the surface gravities of DA1 and DA2, in conjunction with the evolutionary models of \cite{fontaine01}, to estimate their masses and cooling times. For consistency with our previous work and a large number of other recent studies \citep[e.g.][]{dobbie09a,williams09,kalirai08,liebert05a} we have adopted the calculations which include a mixed CO core and thick H surface layer structure.
It is important to note that DA1 has an effective temperature of $T_{\rm eff}$$<$12000K. Spectroscopic mass determinations are known to be systematically larger in this regime than at higher effective temperatures where they agree well with those derived from gravitational redshifts \citep[e.g.][]{bergeron95,reid96}. This trend is most likely due to shortcomings in the treatment of convection within the model atmosphere calculations \citep{koester10}. Based on the studies of \cite{tremblay11}, \cite{koester09} and \cite{kepler07}, we have estimated the size of this effect at $T_{\rm eff}$=10500K to be $\Delta$M=+0.16$\pm$0.04M$_{\odot}$. The properties reported in Table~\ref{temps} reflect a downward adjustment of this size to the estimated mass of DA1.
We have used the grids of synthetic photometry of \cite{bergeron95}, which have been updated by \cite{holberg06}, to derive the absolute $r$ magnitudes  of DA1 and DA2 and to determine the distance moduli of their host binary systems (Table~\ref{temps}). We have neglected foreground extinction in these directions as dust maps suggest A$_{V}$$<$0.1 integrated along these sight lines through the Galaxy \citep{schlegel98}.

To evaluate the masses of DAH1 and DAH2, we first assumed that they reside at the same distance as their non-magnetic companions. We then determined the radius of each HFMWD through scaling it's distance by the square root of the estimated flux ratio, (f/F)$^{0.5}$, where f is the observed flux at the Earth’s surface and F is the flux at the surface of the white dwarf. The surface flux in the SDSS filters \citep[$r$, $i$ and $z$; e.g.][]{fukugita96} for each object was derived from the non-magnetic, pure-H model white dwarf atmospheres of \cite{holberg06}, where cubic splines were used to interpolate between points in this grid. Next we used the mass-radius relations for non-magnetic DA white dwarfs predicted by the evolutionary models of \cite{fontaine01} to obtain estimates of the masses of DAH1 and DAH2 of $M$=0.62$\pm$0.10M$_{\odot}$ and $M$=0.99$\pm$0.05M$_{\odot}$, respectively. Finally, we used the evolutionary models to determine their cooling times to be $\tau$$_{\rm cool}$=726$_{-107}^{+140}$Myr and $\tau$$_{\rm cool}$=321$_{-40}^{+47}$ Myr, respectively. The uncertainties in all derived parameters here were determined using a Monte-Carlo like approach in which we created 25000 realisations of each binary system under the assumption that the adopted errors on the effective temperatures (2.3 per cent), the surface gravities (0.07 dex) and the observed magnitudes (Table~\ref{phot}) of the component stars are normally distributed. We also noted that the SDSS magnitudes (Table~\ref{phot}) have an absolute precision of 2 per cent \citep{adelman_mccarthy08}.

% 2.3-3.3solar for SDSS\,J150746.48+521002.1
% 2.8-4.2solar for SDSS\,J092646.88+132134.5

\section{The progenitors of the HFMWDs}
\label{disc1}

\subsection{In wide double-degenerate systems}

\begin{table*}
\begin{minipage}{160mm}
\begin{center}
\caption{Progenitor masses and system age estimates for the components of the DA+DAH system SDSS\,J150746.48+521002.1  +  SDSS\,J150746.80+520958.0. These are based on three recent estimates of the form of the IFMR, 1. \citet{dobbie06a}, 2. \citet{kalirai08} and 3. \citet{williams09} and stellar lifetimes from the solar metalicity evolutionary models of \citet{girardi00}.}

\begin{tabular}{ccccccc}
\hline
Component & \multicolumn{2}{c}{IFMR 1} & \multicolumn{2}{c}{IFMR 2} &  \multicolumn{2}{c}{IFMR 3} \\

SDSS & $M$$_{\rm init}$ (M$_{\odot}$) & System age (Myr) & $M$$_{\rm init}$ (M$_{\odot}$) & System age (Myr) &  $M$$_{\rm init}$ (M$_{\odot}$) & System age (Myr) \\ \hline

\multicolumn{1}{l}{J150746.48+521002.1} &3.06$^{+0.66}_{-0.58}$ & 599$^{+357}_{-181}$ & 2.76$^{+0.50}_{-0.47}$ & 753$^{+390}_{-217}$ & 2.76$^{+0.37}_{-0.36}$ & 754$^{+259}_{-167}$ \\ \\
\multicolumn{1}{l}{J150746.80+520958.0} & 5.30$^{+0.86}_{-0.75}$ & 426$^{+67}_{-44}$ & 5.50$^{+0.62}_{-0.59}$ & 416$^{+55}_{-38}$ &5.07$^{+0.41}_{-0.42}$ & 437$^{+52}_{-37}$ \\

\hline
\label{progcool1}
\end{tabular}
\end{center}
\end{minipage}
\end{table*}

From their observed positions on the sky and the distance moduli in Table~\ref{temps}, we determine that the components of System 1 and System 2 have minimum separations of a$\sim$650AU and a$\sim$750AU, respectively. Even assuming that the original orbital separations of these binaries were substantially smaller \citep[e.g.][]{valls88}, the Roche lobes of their components have likely always been much larger than the dimensions of an asymptotic giant branch (AGB) star \citep[r$\simless$4-5AU; ][]{ibenlivio93}. Thus this work doubles the number of known wide, magnetic + non-magnetic double-degenerate binaries, where through their large orbital separations the components could be expected to have evolved essentially as single stars.

Intruigingly, in three of these four pairings now known, the non-magnetic component has a substantially greater mass than the value observed at the prominent peak in the field white dwarf mass distribution \citep[e.g. $M$$\sim$0.6M$_{\odot}$; ][]{bergeron92, kepler07}. The intermediate temperatures ($T_{\rm eff}$=10500-16000K) of these three non-magnetic white dwarfs, coupled with their comparatively high masses link their HFMWD companions to an early-type stellar population. For example, DA2 has a mass of $M$$\approx$0.66-0.74M$_{\odot}$ and a corresponding cooling time of $\tau$$\approx$126-171Myr. Assuming it has evolved essentially as a single object from a star with an initial mass of $M_{\rm init}$$\approx$2.3-3.7M$_{\odot}$ \citep[e.g.][]{williams09,kalirai08,dobbie06a}, allowing for a stellar lifetime as predicted by the solar metalicity model grid of \cite{girardi00}, the total age of the binary is likely to be $\tau$$\simless$1150Myr. Thus the formation of DA2 appears to be associated with a star of initial mass $M_{\rm init}$$>$2.2M$_{\odot}$. Following a similar line of reasoning, from the mass and cooling time of DA1 we infer the age of the host system to be $\tau$$\simless$1300Myr which corresponds to the lifetimes of stars with initial masses, $M_{\rm init}$$>$2.1M$_{\odot}$.

Due to our adopted colour selection criteria, our search for wide double-degenerate binaries is sensitive only to relatively recently formed white dwarfs. For example, a 0.6M$_{\odot}$ white dwarf cools to $T_{\rm eff}$$\sim$9000K \citep[$g$-$r$$\approx$0.0; ][]{holberg06} in only 800Myr. Assuming single star evolution, a massive white dwarf companion to a degenerate formed from a sufficiently long lived progenitor ($M_{\rm init}$$\simless$1.6M$_{\odot}$) will generally always have cooled below our photometric colour limits before the latter has formed. Thus the detection of systems where the components have quite different evolutionary timescales is somewhat disfavoured. If HFMWDs are frequently associated with relatively short main sequence lifetimes, we might expect in a ``blue'' colour-selected survey a low probability of finding them paired with white dwarfs which have ``average'' masses ($M$$\sim$0.6M$_{\odot}$). Interestingly, the DA in the fourth pairing which lies within the SDSS DR7 footprint but which failed our survey selection criteria on the grounds of one component being too red (the DAH), has a comparatively low mass, $M$=0.54M$_{\odot}$. The large difference of $\sim$1.6Gyr between the cooling times of PG\,1258+593 and SDSS\,J130033.48+590407.0 means that, within the measurement uncertainties, this HFMWD could still be the progeny an early-type star with $M_{\rm init}$$\approx$2-3M$_{\odot}$ \citep{girven10}.

\begin{figure}
\includegraphics[angle=270,width=8.25cm]{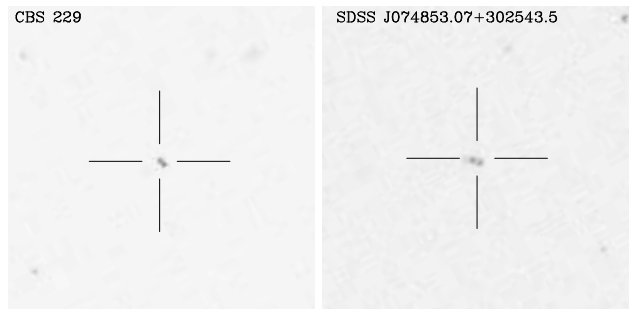}
\caption{SDSS $z$ band image of two candidate spatially resolved magnetic + non-magnetic double-degenerate systems, CBS\,229 \citep{gianninas09} and SDSS\,J074853.07+302543.5. Images are approximately 1'$\times$1' with N at the top and E to the left.}
\label{newcands}
\end{figure}

\subsection{In closer or less well characterised systems}

\begin{table*}
\begin{minipage}{160mm}
\begin{center}
\caption{Progenitor masses and system age estimates for the components of the DAH+DA system  SDSS\,J092647.00+132138.4  +  J092646.88+132134.5. These are based on three recent estimates of the form of the IFMR, 1. \citet{dobbie06a}, 2. \citet{kalirai08} and 3. \citet{williams09} and stellar lifetimes from the solar metalicity evolutionary models of \citet{girardi00}.}

\begin{tabular}{ccccccc}
\hline
Component & \multicolumn{2}{c}{IFMR 1} & \multicolumn{2}{c}{IFMR 2} &  \multicolumn{2}{c}{IFMR 3} \\

SDSS & $M$$_{\rm init}$ (M$_{\odot}$) & System age (Myr) & $M$$_{\rm init}$ (M$_{\odot}$) & System age (Myr) &  $M$$_{\rm init}$ (M$_{\odot}$) & System age (Myr) \\ \hline

\multicolumn{1}{l}{J092647.00+132138.4} & 3.73$^{+0.78}_{-0.69}$ & 1092$^{+202}_{-94}$ & 3.61$^{+0.63}_{-0.65}$ &  1115$^{+178}_{-58}$ & 3.48$^{+0.47}_{-0.51}$ & 1146$^{+134}_{-42}$ \\ \\
\multicolumn{1}{l}{J092646.88+132134.5} & 2.49$^{+0.92}_{-0.85}$ & 1531$^{+1334}_{-372}$ & 2.08$^{+0.96}_{-0.92}$ &  2072$^{+4464}_{-781}$ & 2.18$^{+0.77}_{-0.77}$ & 1892$^{+2215}_{-549}$ \\
\hline
\label{progcool2}
\end{tabular}
\end{center}
\end{minipage}
\end{table*}

Three further hot ($T_{\rm eff}$$\simgreat$9000K) magnetic + non-magnetic binaries have been identified, and spectroscopically confirmed, to date (the DAH component of G\,62-46 has only $T_{\rm eff}$$\sim$6000K). The components of at least one of these systems, LB\,11146 (PG\,0945+245), are separated by only a$\sim$0.6AU \citep{nelan07}. As this is less than the radius of an AGB star, it is likely they have interacted during prior phases of their evolution. Detailed analysis of this binary has revealed it to consist of a $T_{\rm eff}$$\sim$14500K, $M$$\approx$0.9M$_{\odot}$ DA and a similarly hot, massive magnetic white dwarf with a field strength $B$$\simgreat$300MG \citep{liebert93}. 2RE\,J1440+750 (EUVE\,J1439+75.0) was shown, through a detailed spectroscopic (and imaging) analysis, to consist of a $M$$\sim$0.9M$_{\odot}$ DA and a $M$$\sim$1.0M$_{\odot}$ DAH ($B$$\sim$14-16MG), each having $T_{\rm eff}$$>$20000K, and with a projected orbital separation of a$\simless$250AU \citep[][]{vennes99}. CBS\,229 was identified as an unresolved DA+DAH binary during the course of a spectroscopic survey of bright white dwarfs ($V$$\le$17.5) drawn from the catalogue of \cite{mccook99}. \cite{gianninas09} have performed a preliminary analysis of a composite spectrum of this pair and find that the non-magnetic component has $T_{\rm eff}$$\approx$15000K and log $g$$\approx$8.5, corresponding to a mass of $M$$\approx$0.9M$_{\odot}$. The shape of their deblended spectrum of the DAH suggests that the two objects have similar effective temperatures. Our examination of the SDSS imaging reveals that the components are in fact resolved into two photocenters with a projected separation of $\sim$1.3 arcsec (see Figure~\ref{newcands}, left). From the \cite{gianninas09} parameters for the DA and its magnitude from the SDSS $z$ imaging (the band in which the objects are most clearly resolved), we provisionally estimate a distance to this binary of d$\sim$140pc and a projected orbital separation of a$\sim$180AU. Thus CBS\,229 appears to be a wide magnetic + non-magnetic double-degenerate system which escaped detection by our survey. This is probably due to the $u$ band magnitude measurement for the NE component which appears to be anomalous. 

Although it is possible that past mass exchange within at least the first of these binaries has influenced the characteristics of their white dwarfs, as is observed in three out of four of the confirmed wide systems, the non-magnetic components in each of LB\,11146, EUVE\,J1439+75.0 and, provisionally, CBS\,229 also appear to have substantially greater masses than are typical of field degenerates. This is consistent with their magnetic white dwarf companions being related to an early-type stellar population.

\section{Cooling times of the HFMWDs and canonical stellar evolution.}

\subsection{The DAH SDSS\,J150746.80+520958.0}

DA2 has a mass and a cooling time which are comparable to several degenerate members of the Hyades \citep[e.g. WD\,0352+098, WD\,0421+162, ][]{claver01}. This suggests that the host binary system is likely to have a total age which is similar to this cluster \citep[$\tau$=625$\pm$50Myr,][]{perryman98}. We have made detailed estimates of this age using our determinations of the progenitor masses of DA2 and DAH2 (obtained from three recent, independent, derivations of the IFMR), the stellar lifetimes as predicted by the solar metalicity models of \cite{girardi00} and by assuming standard single star evolution (Table~\ref{progcool1}). We find that our estimates are only formally consistent within their quoted (1$\sigma$) error bounds when the oldest, least well constrained of the three approximations to the IFMR \citep{dobbie06a} is adopted. When either of the two other IFMRs is assumed, the age derived from the HFMWD is lower than that obtained from the DA.

This discord is not statistically significant alone but it is notable for being in the same sense as seen for RE\,J0317-853 \citep{ferrario97} and the Praesepe cluster HFMWD, EG\,59 \citep{claver01}. It has been proposed that the age paradox of the RE\,J0317-853  +  LB\,9208 system is due to the former component having formed through the merging of the white dwarf progeny of two stars of more modest initial mass (than is assumed in the case of single star evolution). This could also explain why this HFMWD is observed to rotate with a relatively short period of only 725\,sec \citep{ferrario97}. More recently, \cite{kulebi10} have argued that the cooling time of this HFMWD appears at odds with that of LB\,9208 only because the mass of the former has been slightly underestimated ($\sim$5 per cent) due to evolutionary models neglecting the effects of the magnetic field on the structure of the white dwarf. However, our revised calculations suggest that this earlier conclusion is erroneous and that these structural effects are unable to fully account for the age discprepancy. In \cite{kulebi10}, the tables of \cite{holberg06} were extrapolated to estimate the new cooling ages but effectively considered the wrong radii hence incorrect luminosities for the cooling. Here we instead consider a simple approach using Mestel's equation \citep[see][]{shapiro83} for a half carbon half oxygen white dwarf, which is a good approximation especially before the onset of crystallization (Equation~\ref{mestel}),

\begin{equation}
\tau_{\rm cool} \approx 1.1\times10^7 \ \left(\frac{M}{M_{\odot}} \right)^{5/7}
\left(\frac{L}{L_{\odot}} \right)^{-5/7} \ {\rm years}
\label{mestel}
\end{equation}

where $\tau_{\rm cool}$ is the cooling age, $M$ is the mass of the star in solar masses and $L$ is its luminosity in solar luminosity. Given that luminosity is constant, any difference in mass ($\Delta M$) causes an underestimation of cooling age, linearly (Equation~\ref{linear}).

\begin{equation}
\frac{\Delta \tau_{\rm cool}}{\tau_{\rm cool}}  = \frac{5}{7}\frac{\Delta M}{M}
\label{linear}
\end{equation}

Hence if the mass of RE\,J0317-853 is underestimated by $\sim$ 5  per cent, the underestimation of the cooling age would be only $\sim$ 3.5  per cent, which is far smaller than the observed $\sim$ 30 per cent. 

It should be noted that our calculation is based on the assumption that the radius is inflated by 8-10 per cent. It is possible that the stellar interior contains a greater level of magnetic energy which might have an even stronger influence on the structure, especially if this approaches 10 per cent of the gravitational binding energy as suggested by \cite{ostriker68}. However, this supposition was not based on quantitative considerations of the stellar structure. Recently, \cite{reisenegger09} investigated the magnetic structure of non-barotropic stars and put an upper limit on the internal magnetic energy which can be supported. This value is limited by the entropy of the star. For white dwarfs it is sufficient to consider only the ions hence entropy is directly related to the core temperature. A 1.1$M_{\odot}$ white dwarf with an effective temperature of 18000K can support a magnetic energy at most $\sim$ 2.4  per cent of its gravitational binding energy. In this case the radius is 16 per cent larger, which corresponds to that of a 1.0$M_{\odot}$ non-magnetic white dwarf. Thus the most extreme mass underestimation expected for DAH2 is 10 per cent which translates to a cooling age increase of only $\sim$7 per cent, compared to an observed discrepancy of perhaps $\sim$50 per cent.

\begin{table*}
\begin{minipage}{168mm}
\begin{center}
\caption{Additional parameters of the DA component SDSS\,J150746.48+521002.1 including the radial velocity and the heliocentric space velocity.}
\begin{tabular}{ccccccccccc}
\hline
SDSS & M$_{\rm WD}$ &  R$_{\rm WD}$ & H$\alpha$ shift & $v$ & rv$_{\rm WD}$ & U & V & W \\ 
& \multicolumn{1}{c}{(M$_{\odot}$)} & \multicolumn{1}{c}{(R$_{\odot}$$\times$10$^{-3}$)} & \multicolumn{6}{c}{(kms$^{-1}$)} \\ \hline
\multicolumn{1}{l}{J150746.48+521002.1} & 0.695$\pm$0.043 & 11.9$\pm$0.6 & 19.1$^{+2.8}_{-4.1}$   & 37.1$\pm$4.0 &  -18.0$^{+4.9}_{-5.7}$ & $-$20.6$\pm$3.0 & $-$20.8$\pm$2.8 & $-$6.4$\pm$4.6 \\
\hline
\label{rvfit}
\end{tabular}
\end{center}
\end{minipage}
\end{table*}

\subsection{The DAH SDSS\,J092646.88+132134.5}

We have similarly estimated the progenitor masses of DA1 and DAH1 and calculated the total age of this other new binary, under the assumption of standard single star evolution. These two sets of estimates, which are shown in Table~\ref{progcool2}, are not formally consistent within their 1$\sigma$ errors for any of our three adopted IFMRs. However, as in the case of our other binary system, the discrepancy is not overwhelming, statistically. The variance suggested here is in the opposite sense to that observed for the three HFMWDs discussed previously, with DAH1 appearing too old for its mass or, alternatively, too low in mass for its cooling time. It could be expected, on the basis of single star evolution, that since the white dwarfs in this system have similar cooling times, they should have comparable masses, having descended from stars with similar initial masses and lifetimes. DAH1 may have endured greater mass loss than assumed for a single star. This could be a consequence of having a close companion. Studies of stellar multiplicity have revealed that at least 10 per cent of stars are members of triple or higher order systems \citep{raghavan10,abt76}. For reasons of dynamical stability, these systems are frequently hierarchically structured with triples often consisting of a body in a relatively wide orbit around a much closer pairing \citep{harrington72}. DA1 and DAH1 perhaps trace what was the wider orbit of a putative triple system, with the latter object possibly having been (or perhaps still being) part of a tighter pairing. A $M$$_{\rm init}$$\sim$3.5M$_{\odot}$ star that experiences Roche Lobe overflow around the time of central helium ignition can lead to the formation of a CO white dwarf with a mass towards the lower end of the range estimated for this HFMWD \citep[e.g.][]{iben85}. However, we note that an observed lack of HFMWDs with close detached companions \citep{liebert05c} has been one of the main arguments that these stars are formed through close binary interaction \citep{tout08}. Whether or not this mechanism was required for the manufacture of either of the two HFMWDs identified in this work, the parameters of their  DA companions argue that they are associated with an early type stellar population. 

The sizeable uncertainties associated with the parameters of the HFMWDs highlight the need to expand substantially the sample of those which are members of either nearby star clusters or wide binary systems so that we can begin to firmly identify any trends in their cooling times and inferred progenitor masses, relative to non-magnetic white dwarfs. For now, a relatively straightforward but useful exercise would involve photometrically monitoring these two new systems to search for short period variability that may reveal evidence which more closely links their evolution and that of the RE\,J0317-853 + LB\,9208 system.

\section{The space velocity of SDSS\,J150746.48+521002.1}

\begin{figure}
\includegraphics[angle=270,width=8cm]{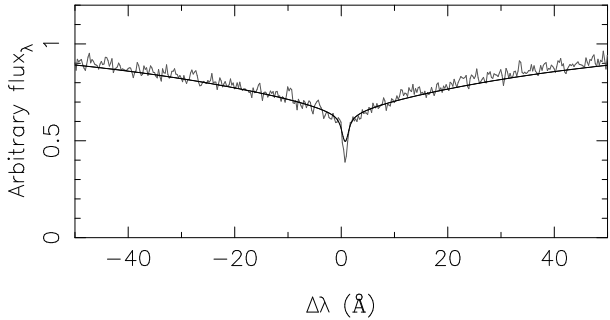}
\caption{The results of our fitting of a synthetic profile (black line) to the central portions of the observed H-$\alpha$ Balmer line of SDSS\,J150746.48+521002.1 (grey lines). The flux$_{\lambda}$ units are arbitrary.}
\label{halp}
\end{figure}

We have exploited the higher resolution spectroscopy we have in hand for DA2 to determine the radial velocity 
of this system from the observed shift of the H$\alpha$ line core. We have removed the effects of telluric water vapour from our red 
arm ISIS data using a template absorption spectrum. The observed H$\alpha$ line and the profile from a non-LTE synthetic spectrum 
corresponding to $T_{\rm eff}$=17500K and log~$g$=8.15, generated using {\tt{TLUSTY}} \citep[v200;][]{hubeny88,hubeny95} and {\tt{SYNSPEC}}
(v49; Hubeny, I. and Lanz, T. 2001, http://nova.astro.umd.edu/), have both been normalised using a custom written IDL routine. The model 
has then been compared to the data, allowing a velocity parameter to vary freely\footnote{An additional flux scaling parameter was allowed to vary 
by up to 1 per cent} and using a Levenberg-Marquardt algorithm to minimise a $\chi$$^{2}$ goodness-of-fit statistic. The result of this process is 
displayed in Figure~\ref{halp} and the line velocity shift, after correction
to the heliocentric rest frame, is shown in Table~\ref{rvfit}. This measurement is not sensitive to the details of the model we adopt for 
effective temperatures and surface gravities within plausible limits. The uncertainty we quote has been estimated using the bootstrapping method of
statistical resampling \citep{efron82}.  Subsequently, the gravitational redshift ($v$ in kms$^{-1}$) component of this velocity shift was derived 
using Equation~\ref{gre}, where $M$ and $R$ are the mass and radius of the white dwarf in solar units respectively.

\begin{equation}
\label{gre}
v = 0.635 M / R
\end{equation}

As described above, the mass and radius of DA2 were determined using the evolutionary tracks of \cite{fontaine01}. 
The radial velocity was then derived from the difference between the measured shift of the line and the calculated gravitational redshift 
(Table~\ref{rvfit}). Finally, adopting the mean of our new (relative)\footnote{The difference between relative and absolute values are 
comparable to or smaller than the quoted errors on the proper motion} proper motion measurements for the components of this binary
 ($\mu_{\alpha}\cos\delta$=30.7$\pm$3.5~mas~yr$^{-1}$, $\mu_{\delta}$=12.9$\pm$4.1~mas~yr$^{-1}$),
we have followed the prescription outlined by \cite{johnson87} to calculate the heliocentric space velocity of DA2 to be
U=$-$20.6$\pm$3.0kms$^{-1}$, V=$-$20.8$\pm$2.8 kms$^{-1}$ and W=$-$6.4$\pm$4.6 kms$^{-1}$. This is coincident with the space velocity of the oldest 
component of the Pleiades moving group, B3, identified by \cite{asiain99} in their Hipparcos kinematic analysis of the B, A and F-type stars in the
 vicinity of the Sun. The sub-populations of this supercluster are estimated to span the range of ages, $\tau$$\approx$60-600Myr \citep{eggen92}. 
While this result is not definitive proof of an association between System 2 and the Pleiades moving 
group, it is at least in accord with our conclusions above, from the cooling time and the inferred main sequence lifetime of the DA component,
that this is a relatively young binary.

\section{Summary and future work}

Within a broader photometric, astrometric and spectroscopic survey for wide double-degenerate systems (Baxter et al. in prep) we have discovered two new binaries, each containing a hydrogen rich HFMWD and a non-magnetic (DA) component. We have used synthetic spectra generated from offset dipole magnetic models to estimate the field strengths for 
DAH1 and DAH2 to be  $B_{\rm dip}$$\sim$210MG ($z_{\rm off}$$\sim$-0.09R$_{\rm WD}$) and $B_{\rm dip}$$\sim$65MG ($z_{\rm off}$$\sim$-0.39R$_{\rm WD}$), respectively. Our measurements of the 
effective temperatures and surface gravities of their non-magnetic companions allow us to infer the masses of these DAHs to be $M$=0.62$\pm$0.10M$_{\odot}$ 
and $M$=0.99$\pm$0.05M$_{\odot}$, respectively. If we assume that the two components in each of these systems have evolved essentially as single stars we find mild 
discord in their cooling times, with DAH2 appearing slightly too hot and young relative to expectations, while DAH1
appears to be ``too old'' for its mass or, alternatively, too low in mass for its cooling time. The former object may represent the third known HFMWD which is apparently ``too young'', perhaps hinting at a trend, though the study of more such systems are required to firm up this possibility. The latter white dwarf may have been part of a hierarchical triple system and suffered greater mass loss than expected of a single star during its earlier evolution. 

In three of the four of these wide systems which are now known, the non-magnetic components have larger masses than are typical of field white dwarfs.
The characteristics (ie. masses, cooling times and kinematics) of the DAs in our two new binaries argue that their HFMWD companions are members of relatively 
young systems and are therefore associated with early type stars ($M_{\rm init}$$\simgreat$2M$_{\odot}$). The non-magnetic components in three additional but spatially unresolved, young, magnetic + non-magnetic
binaries known prior to this work (at least one of which is a physically close system where the components may have previously exchanged mass), also have atypically large masses. 
This is consistent 
with the HFMWDs in these systems also being related to an early type stellar population. To re-inforce these findings and to clearly delineate any trends in the cooling 
times of HFMWDs which could shed light on their formation and the impact of magnetic fields on stellar evolution, an enlarged sample 
of these objects that are located either in wide double-degenerate binaries or in nearby open clusters, will be required. As a starting point we have flagged the previously known DA+DAH system 
CBS\,229 as a probable wide binary. Additionally, we note that the SDSS spectrum of one component of the close ($\sim$1.6 arcsec) pair of relatively bright ($g$$\sim$17.6 mag.),
blue point sources, SDSS\,J074853.07+302543.5 and SDSS\,J074852.95+302543.4 (Figure~\ref{newcands}), displays a Zeeman split, pressure broadened, Balmer line series and thus also represents a promising candidate wide double-degenerate system containing a DAH.

\section*{Acknowledgments}
The WHT is operated on the island of La Palma by the Isaac Newton Group in the Spanish Observatorio del Roque de los Muchachos 
of the Instituto de Astrofísica de Canarias. Funding for the SDSS and SDSS-II has been provided by the Alfred P. Sloan Foundation, the 
Participating Institutions, the National Science Foundation, the U.S. Department of Energy, the National Aeronautics and Space Administration,
 the Japanese Monbukagakusho, the Max Planck Society, and the Higher Education Funding Council for England. The SDSS Web
Site is http://www.sdss.org/. The SDSS is managed by the Astrophysical Research Consortium for the Participating Institutions. The 
Participating Institutions are the American Museum of Natural History, Astrophysical Institute Potsdam, University of Basel, University of 
Cambridge, Case Western Reserve University, University of Chicago, Drexel University, Fermilab, the Institute for Advanced Study, the Japan 
Participation Group, Johns Hopkins University, the Joint Institute for Nuclear Astrophysics, the Kavli Institute for Particle Astrophysics 
and Cosmology, the Korean Scientist Group, the Chinese Academy of Sciences (LAMOST), Los Alamos National Laboratory, the Max-Planck-Institute
for Astronomy (MPIA), the Max-Planck-Institute for Astrophysics (MPA), New Mexico State University, Ohio State University, University of 
Pittsburgh, University of Portsmouth, Princeton University, the United States Naval Observatory, and the University of Washington. BK acknowledges support by the MICINN grant AYA08-1839/ESP, by the ESF EUROCORES Program EuroGENESIS (MICINN grant EUI2009-04170), by the 2009SGR315 of the Generalitat de Catalunya and EU-FEDER funds. NL acknowledges funding from Spanish ministry of science and innovation through the national program AYA2010-19136.

%
%%%%%%%%%%%%%%%%%%%%%%%%%%%%%%%%%%%%%%%%%%
%%%%%%%%  Bibliography  %%%%%%%%
%%%%%%%%%%%%%%%%%%%%%%%%%%%%%%%%%%%%%%%%%%
%
\bibliographystyle{mn2e}
\bibliography{mnemonic,therefs}

\bsp

\label{lastpage}

\end{document}